\documentclass[a4paper,fleqn,usenatbib]{mnras}

\usepackage{mathptmx}

\usepackage[T1]{fontenc}
\usepackage{ae,aecompl}

\usepackage{multirow}     
\usepackage{threeparttable}

\usepackage{graphicx}	
\usepackage{amsmath}	
\usepackage{amssymb}	
\usepackage{subfig}     
\usepackage{bm}		
\usepackage{pdflscape}	

\title[Test of CRF source selection]{Test of source selection for constructing a more stable and uniform celestial reference frame}

\author[N. Liu et al.]{
N. Liu,
J.-C. Liu\thanks{E-mail: jcliu@nju.edu.cn},
Z. Zhu
\\
School of Astronomy and Space Science, Key Laboratory of Modern Astronomy and
Astrophysics (Ministry of Education), Nanjing University,
\\163 XianLin Avenue,
210023 Nanjing, PR China
}

\date{Accepted 2016 November 30. Received 2016 November 30; in original form 2016 November 13}

\pubyear{2016}

\begin{document}
\label{firstpage}
\pagerange{\pageref{firstpage}--\pageref{lastpage}}
\maketitle

\begin{abstract}
We aim to evaluate the possibility of improving the ICRS realization starting
from the ICRF2 catalogue by investigating the coordinate time series of radio sources observed by
VLBI between 1979 and 2016.
Sources with long observational history are selected as the candidates and
the least squares fits with special handling of the weights are performed to derive
the linear drifts of the source coordinates. Then the
sources are sorted based on the normalized linear drift (i) over the whole sky and (ii)
in four homolographic areas divided by declinations. The axial stability of the
reference system and sky distribution defined by the selected sources are
evaluated, which are acted as the criterion for the final source lists.
With our improved source selection scheme, two groups of sources are proposed and
considered suitable for defining a more stable and homogeneous
celestial reference system compared to the current ICRF2. The number of
sources in the final lists are 323 and 294, respectively, and the global rotation of the axes derived
from apparent motion of the sources are about two times better than the ICRF2.
\end{abstract}




\section{Introduction}
In 1994 the International Astronomical Union (IAU) recommended the adoption of
the International Celestial Reference System \citep[ICRS][]{Arias1995}, which is realized
by the highly precise positions of a specific set of extragalactic radio
sources observed with the very long baseline interferometry
(VLBI), known as the International Celestial Reference Frame (ICRF). 
The first version of the ICRF (hereafter ICRF1) was developed by
\citet{ma1998}, based on 212 defining sources with the positional accuracies better
than 1 milli-arcsecond (mas) in both coordinates. 
However, as pointed out by many authors, there are unknown physical characteristics of radio sources,
causing a large drift of coordinates. 
Several subsequent studies were performed \citep[see e.g.][]{Feissel2000,AMGontier2001,FV2003,FV2006,gontier2008,Lambert2009}, assessing the positional stability for individual sources and the celestial frame axes and proposing ensembles of improved source lists.
These ensembles improves the positional stabilities of individual sources and the axial stability of the reference frame.

In 2009 the updated version of the ICRF (hereafter ICRF2) was constructed, which includes 3414 sources and 295 defining sources therein \citep{2009ITN....35....1M,IERS2}. 
The ICRF2 improves the axial stability by a factor of two compared with the ICRF1 and includes more sources in the southern hemisphere, leading to a more uniform sky distribution. 
But the stability estimations of the ICRF2 axes, especially in post-ICRF2 observations, should be continued. 
In a recent paper, \citet{Lambert2013} has proved that there is no significant deformations of the ICRF2 axes by studying the yearly differential reference frames, but the author suggested that such work should be undertaken regularly as the time series update. 
Since the publication of the ICRF2,  time series of new VLBI observations longer than 7 years are available. 
It would be interesting to look into the possibility of upgrading the source selection. 
Since the ICRF3, which is the next generation of standard celestial reference frame in radio wavelength, aims to improve the positional accuracy and sky distribution of sources \citep{2014jsrs.conf...51J,2014AAS...22325125J, 2015jsrs.conf....3M}, we will focus on selecting sources to meet these requirements.

In fact, the selection of sources to construct a reference frame is always complicate and delicate, related to the apparent and intrinsic characteristics of the sources in many aspects. 
Several criteria for choosing sources were adopted in the previous studies. 
Three aspects of radio sources were mainly investigated in the work for ICRF1 \citep{ma1998,arias2004}: (i)quality of data and observational history; 
(ii) consistency of coordinates derived from subsets of data; 
and (iii) repercussions of source structure. 
Using these criteria, sources are categorized into three class: defining, candidate, and other. 
A method of selecting sources based on the analysis of time series stability of astrometric positions was initially proposed by \citet{FV2003} and this work was extended in \citet{FV2004,FV2006}. 
A similar selection scheme can be found in \citet{gontier2008, Lambert2009, le2010time,le2014evaluation}. Several parameters of time series were tested, i.e., standard deviation, slope, Allan standard deviation and goodness of fit, while session time series and regular time series (for example, one-year average) show
different statistical features. 
The selection of the ICRF2 defining sources also partly depends on the time series \citep{2009ITN....35....1M,IERS2} of source coordinates. 
A stability criterion based on overall positional index and successive structure index were applied, according to which the sources were sorted from the most stable to the least. 
To achieve more uniform sky coverage, loose threshold was set for sources in the southern hemisphere.

In this paper, time series of coordinates are used to select suitable sources.
The principle strategy is to obtain new source lists by eliminating unstable sources from and adding new stable sources to the current ICRF2 defining source list. 
The observational history of the sources is considered in Sect.~\ref{sect:preselection}, while a detailed description of further selection schemes is given in Sect.\ref{sect:select}. 
Sect.~\ref{sect:conclusion} presents some discussions and conclusions.

\begin{figure*}
   \centering
   \subfloat[]{\includegraphics[width=\columnwidth]{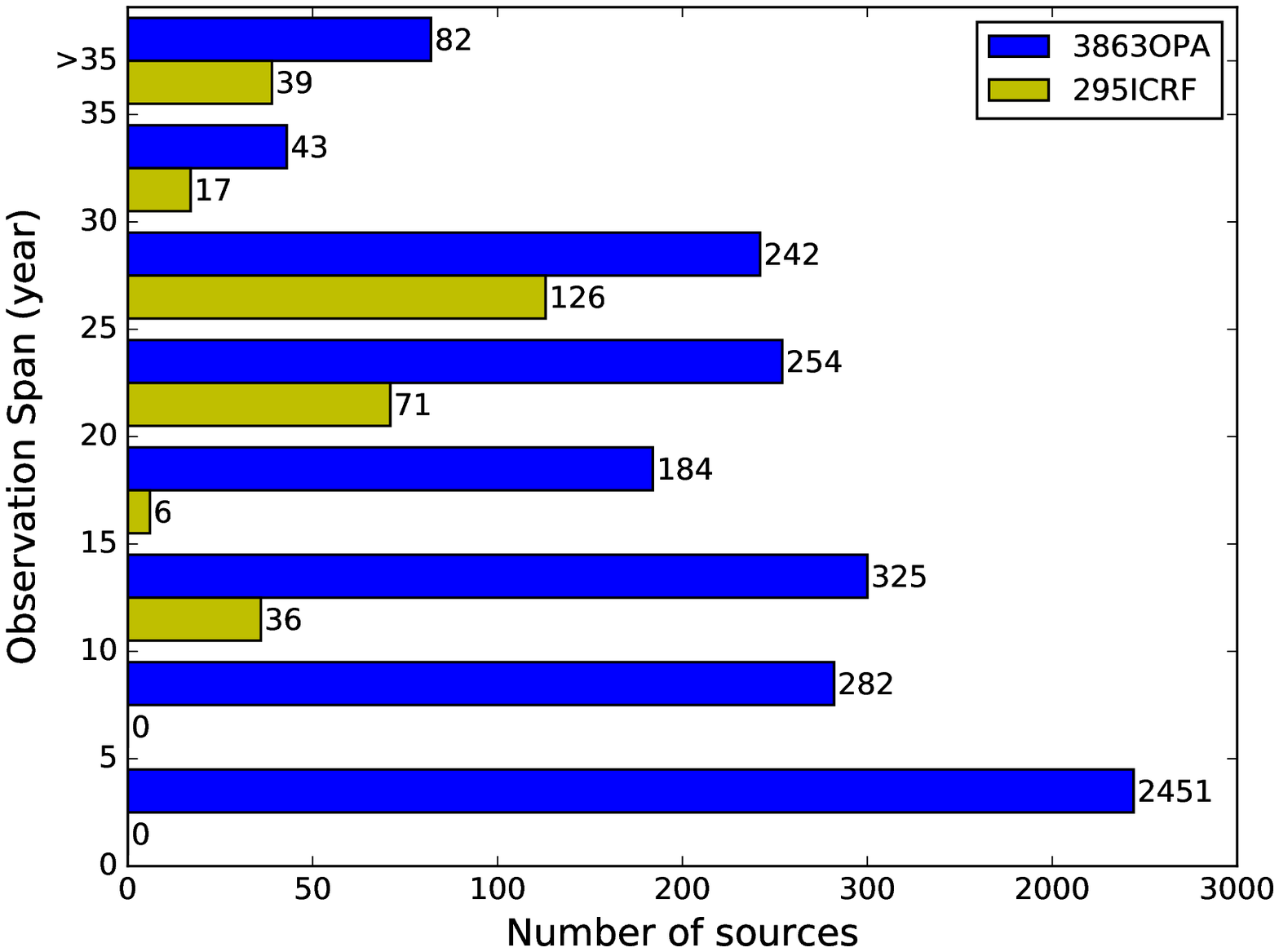}}
   \subfloat[]{\includegraphics[width=\columnwidth]{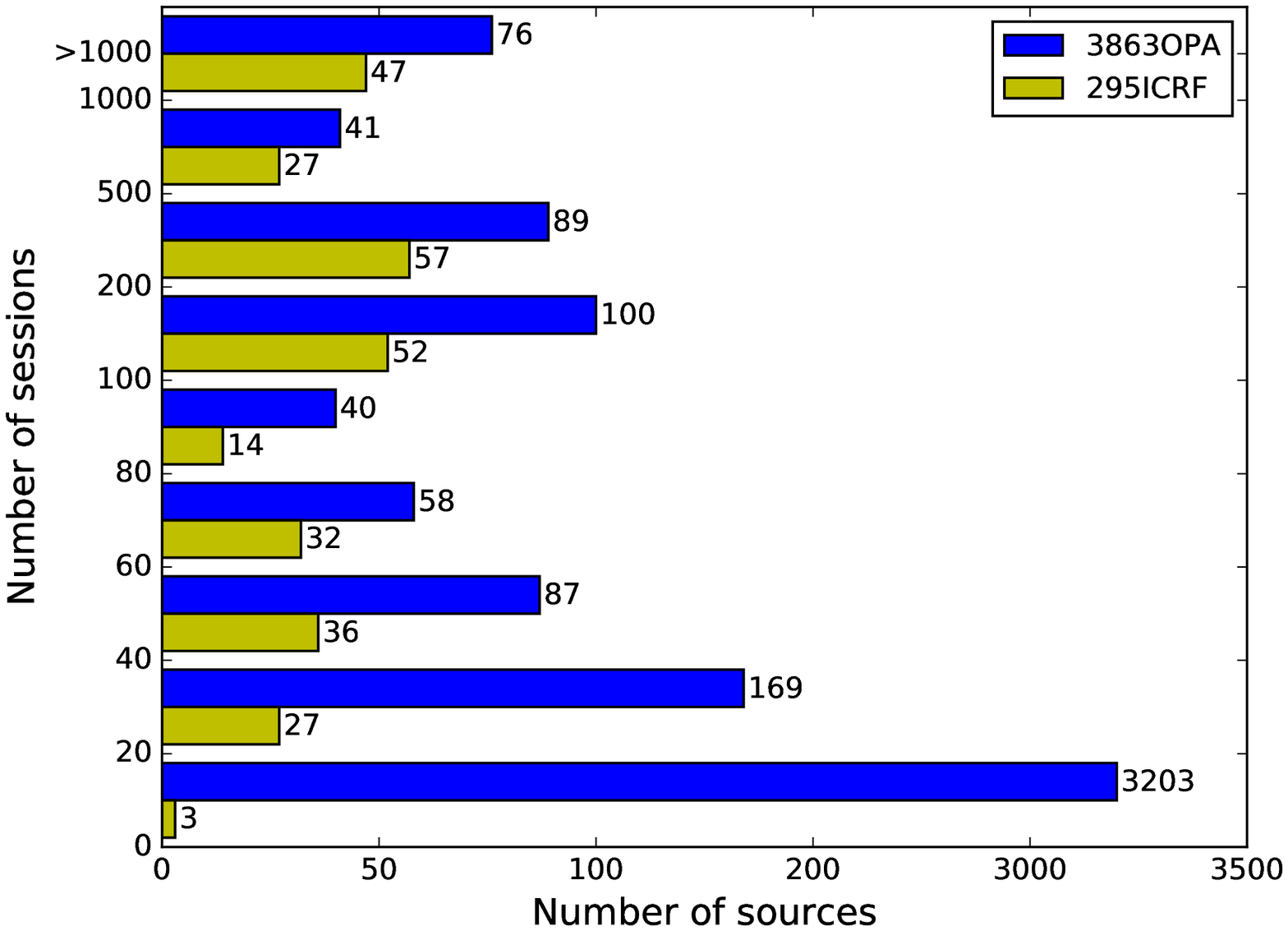}}
\caption{The observational history of the 3863 sources. Panel (a) is the statical histogram of the observation span and panel (b) shows the
amount of the observational sessions for individual sources.
              }
         \label{Fig:Observation}
   \end{figure*}

For comparison, several ensembles of sources proposed in the previous studies
are quoted;
for clarity, in the following sections, the ICRF1 and ICRF2 defining source lists will be referred as ``212ICRF" and ``295ICRF" respectively, while the subset of 247 sources provided by \citet{FV2006} will be denoted as ``247MFV" and the 260 sources proposed by \citet{Lambert2009} as ``260AMS"
(An ensemble of 262 sources was proposed in the paper, but only 260 sources are
contained in the available sources list file).
\section{Data and Pre-selection}\label{sect:preselection}
The data used here are the VLBI derived coordinate time series for 3826
sources provided by IVS analysis center at Paris
Observatory\footnote{\url{http://ivsopar.obspm.fr/radiosources/}}
\citep[see][Sect.~2 for details]{Lambert2013}. Figure \ref{Fig:Observation}
shows the observational history of the total 3863 sources and this list is
labelled as ``OPA3863". 
We note that some non-defining sources have been observed for a longer period and more frequently than some defining sources. 
This motivates us to check whether other well--observed sources can be qualified
for being selected as defining sources.

Previous studies \citep[e.g.][]{AMGontier2001,Lambert2009} mentioned that
the quality and precision of pre-1990 VLBI data seem worse compared to later
observations, therefore the data before 1990.0 should be used with caution. 
For this reason some studies used the coordinate time series only after
1990.0. 
However, the ICRF2 working group \citep{IERS2} claimed that the positions and corresponding uncertainties generated from the entirely available VLBI observations can represent realistically how confident to use these positions in the future. 
For this reason, the full available time series from August 1979 to January 2016 will be used in this work.

To exclude sources with poor observations or questionable behaviors, a pre-selection algorithm is applied. 
First, 39 special handling sources with known significant positional instability, 3 gravitational lenses, and 6 radio stars \citep[see][Sect.~4]{2009ITN....35....1M} are excluded. 
Then the X-band radio structure index provided in the Bordeaux database \citep{Charlot2013} is applied \
as a priori information to reject sources with the structure index larger than 3.

\begin{figure}
   \centering
   \includegraphics[width=\columnwidth]{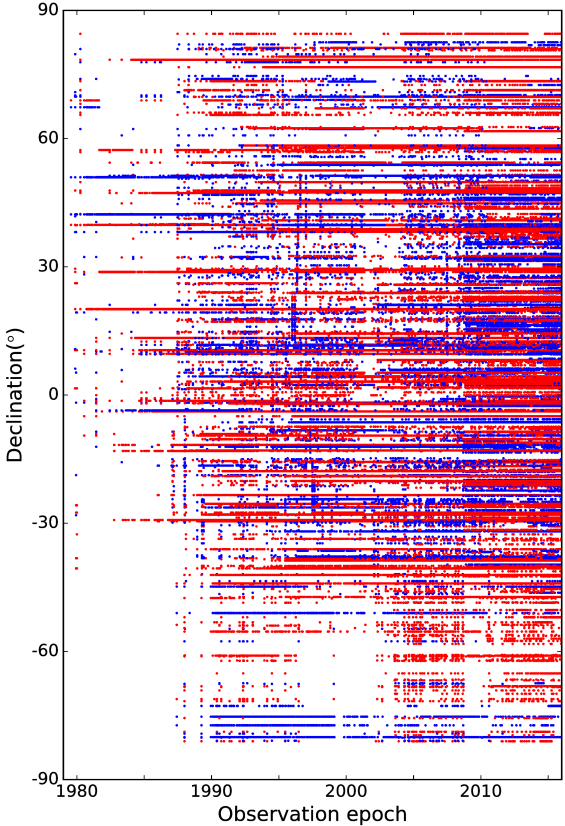}
      \caption{
Observational history of 579 candidates, including 287 ICRF2 defining sources
(red) and 292 non-defining sources (blue). Each circle corresponds to a VLBI session during which the radio source is observed, and radio sources are distinguished by their declinations (the vertical axis.).
              }
         \label{Fig:ObsHis}
\end{figure}

The sources are considered well observed, when the interval of observation is longer than 10 years and number of sessions is larger than 20.
This threshold is artificial since there is no specific definition of a rich or
poor observational history. 
However, taking this filter enables us to keep enough sources for the following studies and eliminate very poor observed sources at the same time. 
It should be noted that the time series for a part of ICRF2 defining sources become obviously denser after 2009, most of them locating in the southern hemisphere. 
These southern sources are kept by our criterion of pre-selection. 
More strict constrains are tested but most of these southern sources will be excluded. 
This is out of our wish since more sources in the southern hemisphere should be added to make the reference frame more uniform. 
Finally, 579 sources including 287 ICRF2 defining sources are retained as candidates
for the next step. 
Eight ICRF2 defining sources (0805+046, 1014+615, 1030+074, 1448-648, 1548+056, 1554-643, 1633-810, 2106-413) are ruled out because of few data points or a large structure index. 
The source names used here are the IERS source designations. 
The observational history of 579 candidate sources is shown in Fig.~\ref{Fig:ObsHis}. 
At low declination zone, several non-defining sources are observed quite frequently, and hence possible to be selected as the defining sources.

\section{Improved Sources Selection Scheme}\label{sect:select}
In principle, the extragalactic radio sources are stationary on the celestial sphere without any transversal velocities (in unit of proper motions) because of their extremely large distances at the level of Mpc. 
However the time series of coordinates still show variability owing to the extended structure, immediate
rejection of jet from the central galaxy, stochastic uncertainties, and systematic errors in the observations. All of these physical and observational effects are reflected in the changes of source positions, which can be described by certain statistical parameters. 
This is the reason why we apply the VLBI time series of source coordinates to estimate individual and global behaviors of selected sources, for the purpose of upgrading the ICRF.

In this work, three parameters deduced from the source coordinates ($\alpha^*, \delta$)\footnote{$\alpha^*=\alpha\cos\delta$}
are calculated: 
(i) the weighted standard deviation referred to the mean (weighted root mean square), 
(ii) the weighted Allan deviation \citep[proposed in][]{Malkin2008}, 
and (iii) the normalized linear drift (the ratio of linear drift to its uncertainty) from the least squares fit. 
To derive the standard deviation and Allan deviation of time series, weighted annual average points are calculated over 1980.0-2016.0. 
The standard deviation describes the scatters of the coordinates while the Allan deviation shows the stochastic properties of the time series and is sensitive to transient abrupt changes of source positions. Combination of these two parameters helps us to eliminate sources with significantly noisy time series.
With enough data points, the fitted linear drift is considered as the indicator for long-term variation.

\begin{figure}
   \centering
   \includegraphics[width=\columnwidth]{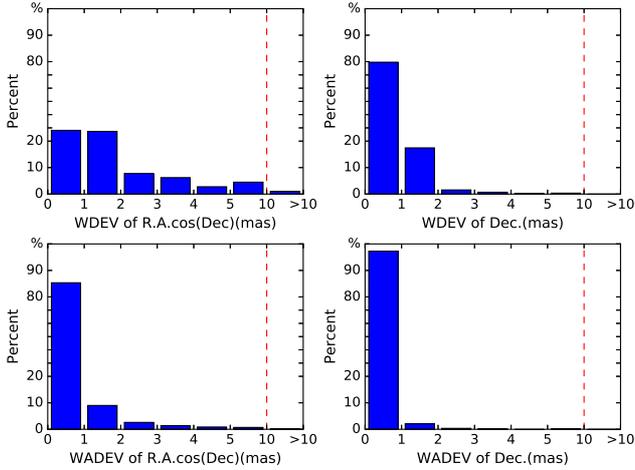}
      \caption{
The statistical histograms of the weighted standard deviation (WDEV) and the
weighted Allan deviation (WADEV) for $\alpha\cos\delta$ ($left$) and $\delta$
($right$) coordinates. The unit is mas. The red vertical lines indicate the upper limit of $10\,\rm{mas}$.
              }
         \label{Fig:Dev}
\end{figure}

Fig.~\ref{Fig:Dev} presents the weighted standard deviation and the weighted Allan deviation for 579 candidate sources.
Our result seems slightly noisier compared to that of \citet{FV2003}. 
Possible reason is that the entire time series are taken into account here while only post-1990.0 time series were used in \citet{FV2003}.
Then the sources with the weighted standard deviation or the weighted Allan
deviation of both coordinates larger than $10\,\rm{mas}$ are excluded.
 As a result, 573 sources are retained.

The linear drifts $(\mu _{\alpha ^*},\mu_\delta)$ are obtained using a weighted least squares fit, but with a
slightly different way of handling weights. 
The time series are divided into three observation spans: 1979.0$\sim$1990.0, 1990.0$\sim$2009.0, and
2009.0$\sim$2016.0 assuming that each observation span corresponds to discrepant accuracy. 
The first interval is considered as it contains relative inaccurate source positions, while the last interval is isolated in order to estimate the effect of post-ICRF2 data. 
The average weight is used as equal weight of session points within the corresponding time span. 
The general linear drift can be written as:
\begin{equation}
\mu = \sqrt{ \mu_{\alpha ^*}^2 + \mu ^2_\delta}.
\end{equation}

\begin{figure}
   \centering
   \includegraphics[width=\columnwidth]{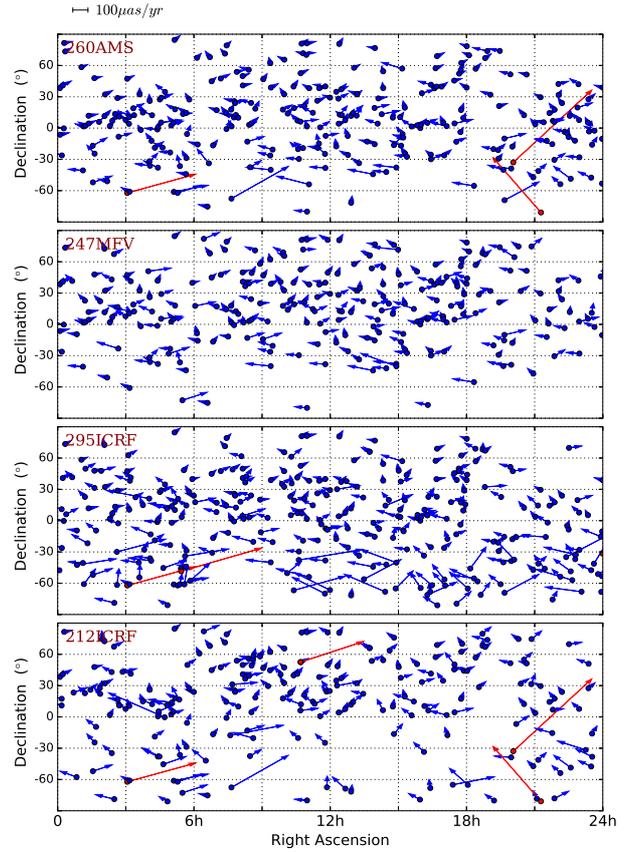}
      \caption{
Linear drift of sources in 212ICRF, 295ICRF, 247MFV and 260AMS source lists. The linear drifts larger than 500 ${\rm \mu
as\,yr^{-1}}$ are distinguished in red.
              }
         \label{Fig:Ld4}
   \end{figure}

In order to set a barrier for extremely large linear drift, we examine the linear drifts of the four source lists provided in literatures, namely 212ICRF, 295ICRF, 247MFV and 260AMS (see Fig.\,\ref{Fig:Ld4}). 
Some sources are found to have very large linear drifts ($>500\,\rm{\mu as\,yr}^{-1}$) and marked in red arrows.
These sources should be eliminated from candidate lists for further steps.
The dimensionless normalized drift ($\mu/\sigma_\mu$ fitted drift divided by its uncertainty) was used in
\citet{Lambert2009} to describe the intrinsic stability of the sources. 
This index is also considered in this study: we keep 565 sources with total linear drift $\mu<500\,\rm{\mu as\,yr}^{-1}$ and sort them from the most stable to the least according to the normalized linear drift $\mu/\sigma_\mu$. 
We call the resulting source list as List1. 
In the next step, we plan to pick sources from the top of the list (with smallest linear drift) to the bottom (with largest linear drift) to form the new defining source list.

Since there is lack of sources and low accuracy in the southern hemisphere, obviously, few
sources will be picked from List1, which would lead to non-uniform distribution on the celestial sphere.
To solve the problem of significant north-south asymmetry in the source numbers, the ICRF2 working group divided the celestial sphere by four nodes of declination into five intervals with approximately the same number of sources, and then applied a loose threshold for low declination sources.
In this paper the sphere is divided into four intervals with the equal spherical area, and the corresponding nodes of the declination are $-30^{\circ}$, $0^{\circ}$, and $+30^{\circ}$.
The numbers of sources locating in the four sub-areas are 99, 128, 176, and 162 (565 in total)
from south to north respectively.
In each belt-like area, sources are sorted according to the normalized linear
drifts as we did for List1 in the whole celestial sphere.
Each time we pick one source from each sub-area (totally four) based on the sequence to form the new list. The resulting new source list is named as List2.
In this way, the numbers of selected defining sources from each sub-area are equal before sources in the smallest sub-group (at the south pole area) are used up.
This approach is applied to balance the requirements of source stability and uniformity of
distribution, which is also considered by ICRF3 working group \citep{2014jsrs.conf...51J,2014AAS...22325125J, 2015jsrs.conf....3M}.
In the next section, we evaluate the property of the reference system realized by our selected sources in global sense and determine the best available source lists.

\subsection{Considerations on the axial stability}
The stability or inertia of a reference frame axes is usually assessed by the
amplitude of global rotation vector $\bm{r}=(r_1,r_2,r_3)^{\rm T}$, where
$r_1$, $r_2$, and $r_3$ are derived by the least squares fit of the linear
drifts to the following equations:
\begin{equation}\label{eq:rotation}
   \begin{array}{lll}
\mu_{\alpha ^*} & = & +r_1\cos\alpha\sin\delta + r_2\sin\alpha\sin\delta -
r_3\cos\delta \\
      \mu_{\delta} 	& = &	-r_1\sin\alpha	 		+ r_2\cos\alpha.
   \end{array}
\end{equation}
Note that some additional parameters such as slopes in right ascension and declination (${\rm d}z$) are occasionally estimated simultaneously \citep[e.g.][]{Lambert2013} for specific purposes, we only take global rotation into account as it is sufficient to estimate the stability of the reference frame.

\begin{figure}
   \centering
   \includegraphics[width=\columnwidth]{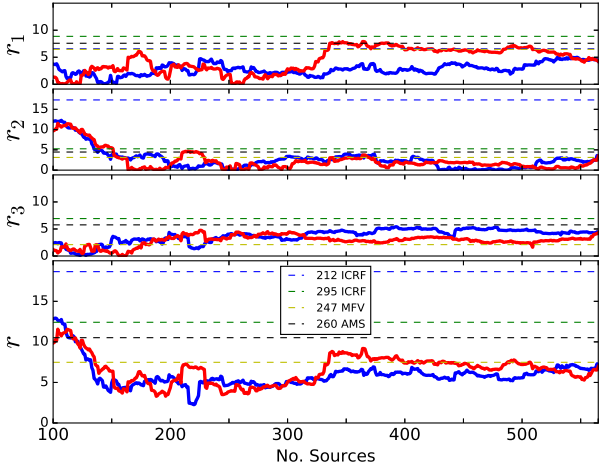}
      \caption{
The fitted rotation vectors correspond to source List1 (blue) and List2
(red). The four horizontal lines are the results of the four special
subsets, as given in the legend.
              }
         \label{Fig:rot_num}
\end{figure}

\autoref{Fig:rot_num} shows the evolution of $r_1, r_2, r_3$ and $r = |\bm r|$ with the number of picked sources from List1 and List2, respectively. 
It can be obviously seen from the trend that as the number increases, more sources with large linear drift $\mu$ are included, causing more significant axial rotations. 
Eclipses and peaks in the curve are also visible. 
For the two original lists, $r$ is approximately equal. 
And the magnitude of $r$ is smaller than that for 295ICRF by a factor of two in some subsets.

In fact, the first order vector spherical harmonics should include glide pattern besides global rotation \citep{Mignard2012}. 
The glide vector $\bm{d}=(d_1,d_2,d_3)^{\rm T}$ is estimated together with $\bm r$ using the following equations:
\begin{equation}\label{eq:rotation&glide}
   \begin{array}{lll}
\mu_{\alpha ^*} 	& = 	& -d_1\sin\alpha + d_2\cos\alpha                      \\&  	& +r_1\cos\alpha\sin\delta +r_2\sin\alpha\sin\delta -r_3\cos\delta   \\\mu_{\delta} & = & -d_1\cos\alpha\sin\delta -d_2\sin\alpha\sin\delta
+d_3\cos\delta \\ & & -r_1\sin\alpha + r_2\cos\alpha.
     \end{array}
\end{equation}

\begin{figure}
   \centering
   \includegraphics[width=\columnwidth]{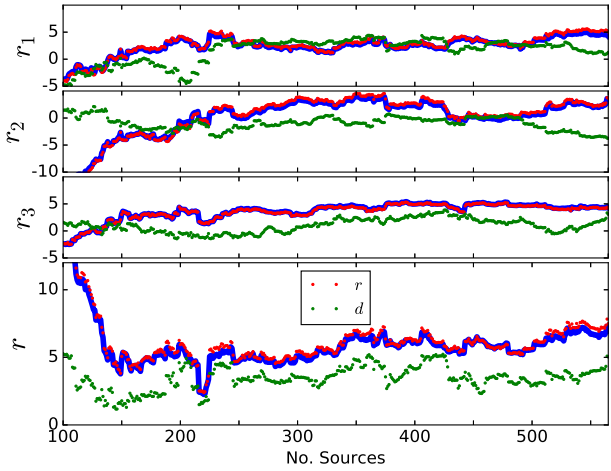}
      \caption{
The fitted rotation and glide vector for source List1.
The blue line is the result shown in Fig. \ref{Fig:rot_num}. The red and green
circles are rotation and glide part respectively. The plots
for List2 are similar.}
      \label{Fig:Rot_Gli}
   \end{figure}

\autoref{Fig:Rot_Gli} displays the values for the components and amplitudes of $\bm r$ and $\bm d$ as the number of selected sources increases. 
It can be easily noticed that the rotation part is close to the one obtained by fitting to Eq.~(\ref{eq:rotation})(result given in  Fig.~\ref{Fig:rot_num}), while the magnitude of glide keeps nearly unchanged.
Therefore it is reasonable for us to consider only rotation part. 
Higher orders of harmonics are not considered due to the smallness of the coefficients and insufficient source numbers.

\subsection{Considerations of the sky distribution}
The level of uniformity for the sources distribution on the celestial sphere is another aspect that needs to be assessed.
The mean declination of subset is one of assessing indexes for such purposes. 
\citet{Liu2012} provided an approximate approach to assess uniformity of source distribution.
In that method, a dipolar vector field is generated based on the coordinates of sources with certain amplitude (e.g $5~{\rm \mu as\,yr^{-1}}$), then global rotation vector  $\bm g$ is obtained with an unweighted least square fit. 
The amplitude of $\bm g$ was proved to be appropriate for describing the uniformity of source distribution on the celestial sphere and used as the homogeneity index, denoted $g$. 
In the present work, the simulation is applied but the direction of the dipolar vector field is set to be the North celestial pole instead of the Galactic center.

\begin{figure}
   \centering
   \includegraphics[width=\columnwidth]{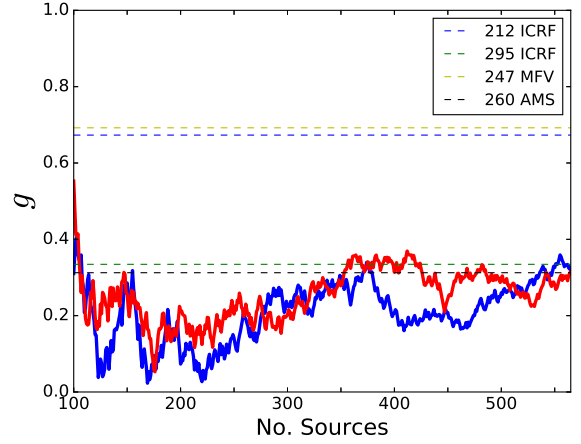}
      \caption{
Results of $g$ for evaluation of source distribution. The blue and red lines correspond to our selected source List1 and List2 respectively.
              }
         \label{Fig:sim}
   \end{figure}

\autoref{Fig:sim} shows the result for $g$ which indicates the level of homogeneity for different source distributions: smaller value for $g$ infers that the distribution is more uniform.
Four horizontal dash lines are the results of 212ICRF, 295ICRF, 247MFV and 260AMS, showing that the 295ICRF and 260AMS have much more uniform source distribution.
For List1 (blue line) and List2 (red line), homogeneity index $g$ ascends when the number of source ensemble increases, despite some fluctuations. 
When the number exceeds around 400, the distribution assessment of source ensemble is worse than those of 295ICRF and 260AMS. 
Recalling that the minimum number of sources for four sub-areas is 99 at the south pole region, north-south asymmetry in the source numbers becomes significant as the total source number reaches 400.

\subsection{Final source lists}
\begin{figure}
   \centering
   \includegraphics[width=\columnwidth]{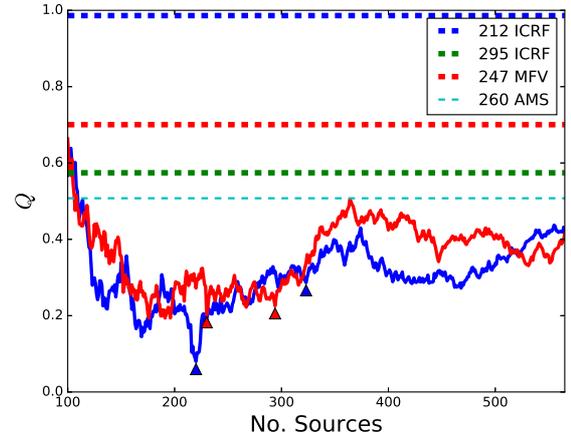}
      \caption{
The Quality index for different source sets when the number increases. 
The blue and red lines correspond to our selected source List1 and List2 respectively.
The final selections are labeled with triangles(Sou220, Sou230, Sou294 and Sou323, from left
to right respectively.)
              }\label{Fig:com_num}
   \end{figure}

When considering the axial stability and sky distribution simultaneously, a quality index of selected source list is defined as
\begin{equation}
Q = \frac{r_N+g_N}{2},
\end{equation}
where $r_N$ and $g_N$ are the results of Fig.~\ref{Fig:rot_num} and
Fig.~\ref{Fig:sim} normalized to unit. 
This weight ratio 1:1 shows a balanced consideration between axial ability and distribution. 
The result of $Q$ for different selections is given in Fig.~\ref{Fig:com_num}. 
The horizontal dash lines represent the parameter $Q$ for existing source lists. 
Finally four ensembles of sources are selected as better representation of the celestial reference system, which are called ``Sou220", ``Sou323" (blue triangles in Fig.~\ref{Fig:com_num}), ``Sou230", and ``Sou294"(red triangles in Fig.~\ref{Fig:com_num}).

\begin{figure}
   \centering
	 \includegraphics[width=\columnwidth]{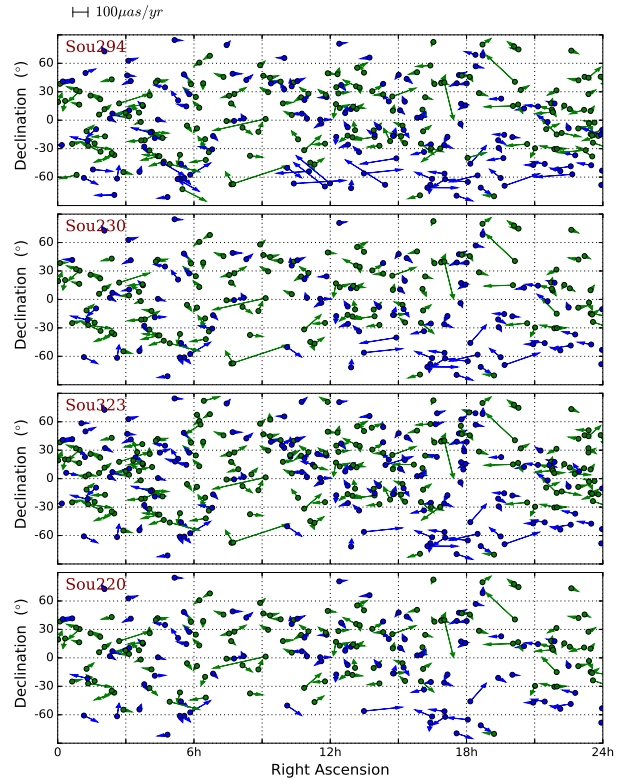}
      \caption{
Linear drifts of sources in Sou220, Sou323, Sou230, and Sou294 lists to be chose for analysis. 
The blue circles indicate the sources included in ICRF2 defining source list.
              }
         \label{Fig:4set}
   \end{figure}

The linear drifts of sources in these four sets are shown in Fig.~\ref{Fig:4set},showing insignificant linear drifts for most sources (the arrows in the figure) and a good sky coverage. 
The blue circles stand for the 295ICRF defining sources while
the green circles represent the non-defining ones added via our selection method. 
Compared with 295ICRF, nearly half sources are removed from the list, which is similar to the result in \citet{le2010time}.

The mean declination and homogeneity index $g$ of the various ensembles are given in \autoref{Tab:Mean_Dec}. 
Comparing the mean declinations of the ensembles from List1 (Sou220 and Sou323) with that from List2 (Sou230 and Sou294), our selection method (dividing the sources into sub-groups according to the
declination) works, leading to a more balanced source distribution about the equator. 
\autoref{Tab:Axi} reports the global rotations derived by the least squares fits for various source lists. Rotations around each axis are around $2\,\mu \rm{as \,yr}^{-1} $,  which are comparable to or smaller than those of the 212ICRF , 295ICRF, MFV247 and AMS260, especially for $r_3$, showing that our selected sources would improve the axial stability of reference frame.
 
\begin{figure*}
   \centering
   \subfloat[Sou323]{\includegraphics[width=\columnwidth]{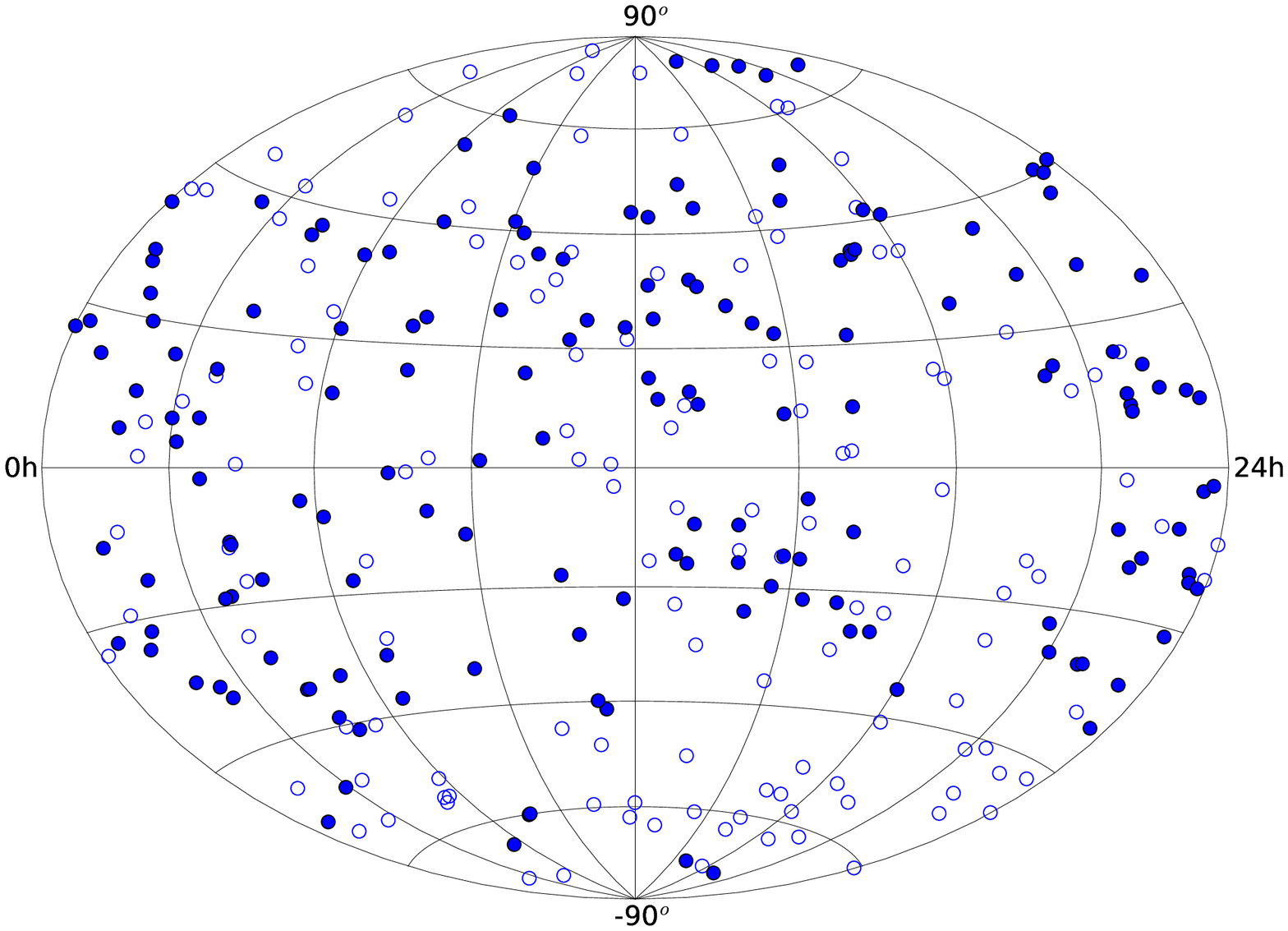}}
   \subfloat[Sou294]{\includegraphics[width=\columnwidth]{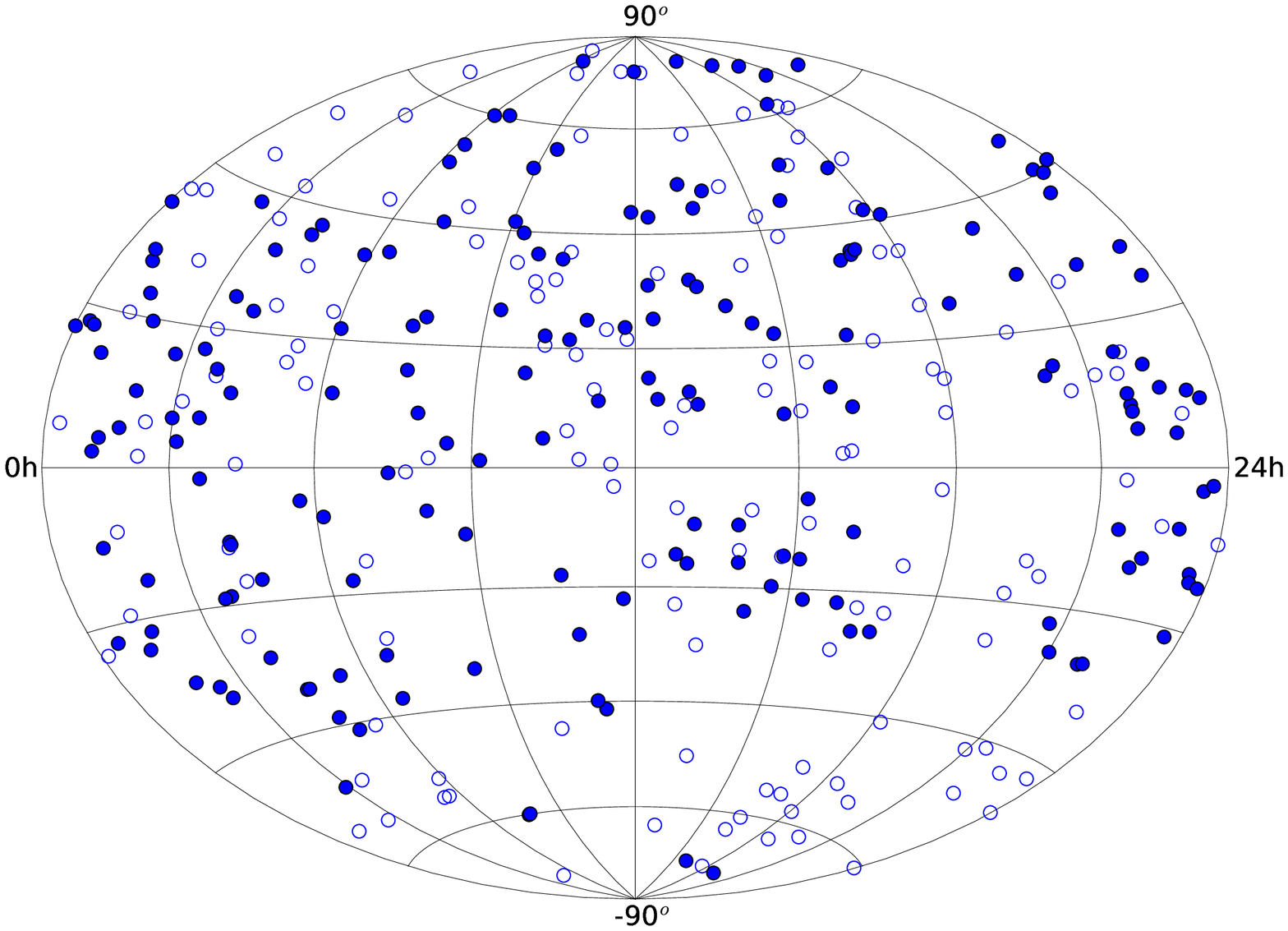}}
\caption{Sky distributions of final source sets Sou323 and Sou294.
The original ICRF2 defining sources are marked by open circles while additional suitable sources identified in this work are plotted as solid circles.
              }
         \label{Fig:SkyDis}
   \end{figure*}

Finally, two source sets, namely Sou323 and Sou294, are proposed as final lists when taking into consideration the distribution of source and stability of the reference axes. 
The sky distribution plots for these two source sets are given in Fig.~\ref{Fig:SkyDis}; source lists and other detailed informations can be found in Table~\ref{tab:sou323} and \ref{tab:sou294}.


	\begin{table}
\caption{Mean declination and homogeneity index $g$ of various source
ensembles.}
	\label{Tab:Mean_Dec}
	\centering
	\begin{tabular}{l r c r}
	\hline\hline
Source list  	& Mean Dec.       	&$g$     		& Number of       \\
    			& ($^{\circ}$)  		&     			&295ICRF  sources \\
\hline
212ICRF  & 14.80		   			&0.67 	&   97                    \\
295ICRF  &   0.70		   			&0.33 	& 295                    \\
247MFV   & 16.89		 			&0.69 	& 133                    \\
260AMS   &   7.97		 			&0.31 	& 148                    \\
Sou220    &   6.13		 			&0.03 	&    93                    \\
Sou323    &   6.84		 			&0.20 	& 141                    \\
Sou230    &  $-$0.64			  		&0.12 	&   99                    \\
Sou294    &  $-$0.98			   		&0.35 	& 130                    \\
\hline
	\end{tabular}
	\end{table}

	\begin{table}
	\centering
\caption{Fitted global rotation for our selected source lists. The unit is $\mu
\rm{as \, yr}^{-1} $.}
	\label{Tab:Axi}	
	\begin{tabular}{ccccc}
	\hline\hline
Sources list      	&$r_1$          		&$r_2$          		&$r_3$          		&$r$    \\
\hline
Sou220 			&+1.7$\pm$4.3 	&$-$0.4$\pm$4.1  	&+1.4$\pm$4.1 	&2.3$\pm$7.2 \\
Sou323 			&+1.3$\pm$3.5 	&+2.7$\pm$3.4 	&+4.4$\pm$3.4 	&5.3$\pm$5.9 \\
Sou230 			&+2.9$\pm$4.6 	&+1.8$\pm$4.5 	&+2.7$\pm$4.5 	&4.4$\pm$7.9 \\
Sou294 			&+1.6$\pm$4.4 	&$-$0.4$\pm$4.2  	&+3.9$\pm$4.4 	&4.3$\pm$7.5 \\
\hline
	\end{tabular}
	\end{table}

\begin{table*}
\caption[]{
Time series statistical information of sources in Sou323. 
For a detailed explanation of columns see Sect.~\ref{sect:preselection} and \ref{sect:select}. Only the first 10 rows are shown here; the full table is available online.
 }
 \label{tab:sou323}
\centering
\begin{threeparttable}
\begin{tabular}{ c c c r r c r r c r r c r }
\hline
\hline
\multirow{2}{*}{Designation \tnote{a}}	
&\multirow{2}{*}{Source\tnote{b} }	
&\multirow{2}{*}{$\mu/\sigma_\mu$}	 &\multicolumn{2}{c }{Linear drift}		&\multirow{2}{*}{}	&\multicolumn{2}{c}{WDEV}	&\multirow{2}{*}{}	
&\multicolumn{2}{c}{WADEV}	
&\multirow{2}{*}{$T_{obs}$\tnote{c}}	
&\multirow{2}{*}{$N_{ses}$}	  \\
\cline{4-5} \cline{7-8} \cline{10-11}
			&		&		&$\mu_{\alpha ^*}$		&$\mu_\delta$		&		&$\alpha ^*$	&$\delta$		&		&$\alpha ^*$	&$\delta$		&		&	\\
			&		&		&$\mu \rm{as \,yr}^{-1}$	&$\mu \rm{as \,yr}^{-1}$	&		&mas	&mas		&		&mas		&mas		&year	&	\\
\hline
ICRF J163231.9+823216  &1637+826  &  0.09  &  -1.0  &  -0.3  &  &  1.64  &  0.34  &  &  0.44  &  0.06  & 18.95  & 207\\
ICRF J221810.9+152035  &2215+150  &  0.09  &  -1.6  &  -0.6  &  &  0.90  &  0.41  &  &  0.25  &  0.11  & 11.20  & 121\\
ICRF J223236.4+114350  &2230+114  &  0.10  &  -0.8  &   0.8  &  &  1.24  &  0.65  &  &  0.33  &  0.17  & 31.55  & 320\\
ICRF J231147.4+454356  &2309+454  &  0.12  &  -1.5  &   2.0  &  &  1.59  &  0.27  &  &  0.91  &  0.14  & 20.11  & 222\\
ICRF J055217.9+375425  &0548+378  &  0.13  &  21.8  &  -5.7  &  &  0.70  &  2.80  &  &  0.15  &  1.28  & 20.27  &  65\\
ICRF J033553.9$-$543025  &0334$-$546  &  0.17  &  12.8  &   2.8  &  &  1.73  &  1.28  &  &  0.56  &  0.39  & 27.79  &  29\\
ICRF J161042.0+241449  &1608+243  &  0.17  &  -3.0  &   6.2  &  &  2.09  &  1.18  &  &  2.29  &  1.34  & 20.37  &  60\\
ICRF J141946.6+382148  &1417+385  &  0.19  &  -1.1  &  -0.7  &  &  0.83  &  0.30  &  &  0.12  &  0.06  & 21.48  & 553\\
ICRF J005655.2+162513  &0054+161  &  0.19  &   3.3  &  -7.7  &  &  0.28  &  0.45  &  &  0.12  &  0.09  & 20.26  & 120\\
ICRF J030903.6+102916  &0306+102  &  0.21  &  -1.3  &  -4.1  &  &  0.36  &  0.64  &  &  0.18  &  0.31  & 27.15  &  99\\
\hline
\end{tabular}
\begin{tablenotes}
        \footnotesize
        \item[a] ICRF Designations.
        \item[b] IERS Designations.
        \item[c] Observation span.
      \end{tablenotes}
    \end{threeparttable}
\end{table*}

\begin{table*}
\caption[]{
Time series statistical information of  sources in Sou294. The data format is the same with \autoref{tab:sou323}. For a detailed explanation of columns see Sect.~\ref{sect:preselection} and \ref{sect:select}. Only the first 10 rows are shown here; the full table is available online.}
\label{tab:sou294}
\centering
\begin{tabular}{ c c c r r c r r c r r c r }
\hline
\hline
\multirow{2}{*}{Designation}	&\multirow{2}{*}{Source}	&\multirow{2}{*}{$\mu/\sigma_\mu$}	&\multicolumn{2}{c }{Linear drift}		&\multirow{2}{*}{}	&\multicolumn{2}{c}{WDEV}	&\multirow{2}{*}{}	&\multicolumn{2}{c}{WADEV}	&\multirow{2}{*}{$T_{obs}$}	&\multirow{2}{*}{$N_{ses}$}	  \\
\cline{4-5} \cline{7-8} \cline{10-11}
			&		&		&$\mu_{\alpha ^*}$		&$\mu_\delta$		&		&$\alpha ^*$	&$\delta$		&		&$\alpha ^*$	&$\delta$		&		&	\\
			&		&		&$\mu \rm{as \,yr}^{-1}$	&$\mu \rm{as \,yr}^{-1}$	&		&mas	&mas		&		&mas		&mas		&year	&	\\
\hline
ICRF J163231.9+823216  &1637+826  &  0.09  &  -1.0  &  -0.3  &  &  1.64  &  0.34  &  &  0.44  &  0.06  & 18.95  & 207\\
ICRF J221810.9+152035  &2215+150  &  0.09  &  -1.6  &  -0.6  &  &  0.90  &  0.41  &  &  0.25  &  0.11  & 11.20  & 121\\
ICRF J033553.9$-$543025  &0334$-$546  &  0.17  &  12.8  &   2.8  &  &  1.73  &  1.28  &  &  0.56  &  0.39  & 27.79  &  29\\
ICRF J232747.9$-$144755  &2325$-$150  &  0.30  &-121.0  & 449.5  &  &  1.93  &  1.28  &  &  1.94  &  2.86  & 23.64  &  27\\
ICRF J223236.4+114350  &2230+114  &  0.10  &  -0.8  &   0.8  &  &  1.24  &  0.65  &  &  0.33  &  0.17  & 31.55  & 320\\
ICRF J231147.4+454356  &2309+454  &  0.12  &  -1.5  &   2.0  &  &  1.59  &  0.27  &  &  0.91  &  0.14  & 20.11  & 222\\
ICRF J053435.7$-$610607  &0534$-$611  &  0.22  & -27.2  &  16.9  &  &  2.24  &  0.79  &  &  0.84  &  0.29  & 23.34  &  31\\
ICRF J060759.6$-$083449  &0605$-$085  &  0.33  &  -7.3  &  -3.8  &  &  0.42  &  0.60  &  &  0.05  &  0.07  & 34.90  &  45\\
ICRF J055217.9+375425  &0548+378  &  0.13  &  21.8  &  -5.7  &  &  0.70  &  2.80  &  &  0.15  &  1.28  & 20.27  &  65\\
ICRF J161042.0+241449  &1608+243  &  0.17  &  -3.0  &   6.2  &  &  2.09  &  1.18  &  &  2.29  &  1.34  & 20.37  &  60\\
\hline
\end{tabular}
\end{table*}

\section{Discussions and conclusions}\label{sect:conclusion}
\subsection{Notes on special handeling sources}
\begin{figure}
   \centering
   \includegraphics[width=\columnwidth]{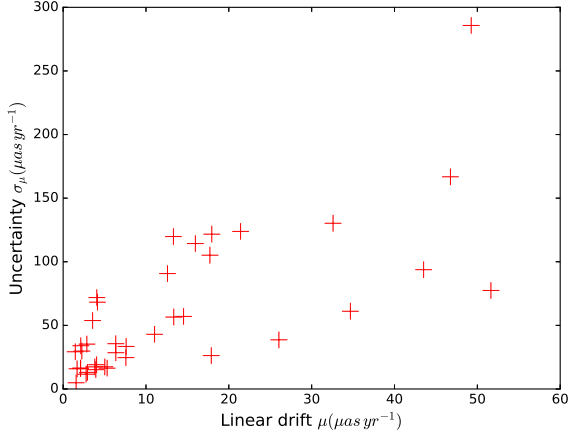}
      \caption{\label{Fig:special_common}
The linear drifts and theirs uncertainties of 39 special handling sources (red plus).
              }
   \end{figure}

\autoref{Fig:special_common} plots the linear drifts and their uncertainties for the 39 special handling sources (red plus), showing that some problematic sources with small normalized linear drift will possibly be chosen as defining. 
When applying our selection scheme to the sources with known poor stability, analysis based only on coordinates time series is not sufficient to detect unstable sources. 
Therefore other information such as the VLBA map is necessarily used together with the coordinates time series for analyzing the stability for individual sources.
So in this study, a pre-selection is applied. 

\subsection{Concluding remarks}   
With the coordinate time series over 30 years, some of ICRF2 defining appear to be unstable and not suitable for defining a celestial frame.
We have applied an improved selection scheme to pick out suitable sources as candidates. 
The scheme is based on observational history and statistical parameters of coordinates time series (weighted standard deviation and weighted Allan deviation) for individual sources. 
Two ways of ranking sources from the most to the least stable are used to obtain source rank lists (List1 and List2).
In both lists, only about half of 295ICRF defining sources are kept (\autoref{Tab:Mean_Dec}). 
The positional differences are represented by the linear drifts in both coordinates and hence the global
rotations for axes of the celestial frame realized by different source sets are estimated, showing that the axial stability is improved by a factor of two with the defining sources selected in this work. 
Moreover, a possible method of estimating the homogeneity of source distribution are used besides the mean declination comparison, which is also the consideration of the future ICRF3 \citep{2014jsrs.conf...51J,2014AAS...22325125J, 2015jsrs.conf....3M}. 
A quality index considering both axial stability and uniform sky distribution is introduced to evaluate the quality of source sets. 
Finally, two sets of sources that show improved axial stability and sky distribution are recommended (Sou323 and Sou294). 

\section*{Acknowledgements}

This research has made use of material from the Bordeaux VLBI Image Database
(BVID) and the IVS analysis center at Paris Observatory. This work is funded by the National
Natural Science Foundation of China (NSFC) under grant Nos. 11303018 and
11473013 and the Natural Science Foundation of Jiangsu Province under No.
BK20130546. We thank Dr. D.-L Liu and our colleague L.-W.-J Zhou for their careful corrections of the English. We are indebted to the anonymous referee for his/her constructive comments on the original version of the manuscript.




\bibliographystyle{mnras}
\bibliography{SourcesSelection} 






\bsp	
\label{lastpage}
\end{document}